\documentclass[a4paper]{llncs}
\pdfoutput=1

\usepackage{graphicx}
\usepackage[utf8]{inputenc}
\usepackage{url}
\usepackage{wrapfig}

\title{Online open neuroimaging mass meta-analysis}
\author{Finn Årup Nielsen\inst{1}\thanks{Thanks to the Lundbeck
    Foundation for the funding of the \emph{Center for Integrated Molecular Brain
    Imaging} (CIMBI).} 
  \and Matthew J. Kempton\inst{2} 
  \and Steven C. R. Williams\inst{2}} 

\authorrunning{Nielsen, Kempton, Williams}
\institute{DTU Informatics, Technical University of Denmark, Lyngby,
  Denmark. 
  \email{fn@imm.dtu.dk}, \texttt{http://www.imm.dtu.dk/\~{}fn/}
\and
Department of Neuroimaging, Institute of Psychiatry, King's College
London, London, UK} 

\begin{document}
\maketitle
\setcounter{footnote}{0}

\begin{abstract}
  We describe a system for meta-analysis where a wiki stores numerical
  data in a simple format and a web service performs the numerical
  computation.  
  We initially apply the system on multiple meta-analyses of
  structural neuroimaging data results. The described system allows for
  mass meta-analysis, e.g., meta-analysis across multiple brain
  regions and multiple mental disorders.
\end{abstract}

\section{Introduction}

The scientific process aggregates a large number of scientific results
into a common scientific consensus.   
\emph{Meta-analysis} performs the aggregation by statistical analysis of
numerical values presented across scientific papers.
Collaborative systems such as wikis may easily aggregate text and
values from multiple sources.
However, so far they have had limited ability to apply
numerical analysis as required, e.g., by meta-analysis.

Researchers have discussed the advantages and disadvantage of the
tools for conducting systematic reviews from ``paper and pencil'', 
over spreadsheets to RevMan and web-based specialized applications
\cite{ElaminM2009Choice}: Setup cost, versatility, ability to manage
data, etc. 
In 2009 they concluded that ``no single data-extraction method is best
for all systematic reviews in all circumstances''.
For example, RevMan and Archie of the Cochrane Library provide an elaborate system for
keeping track and analyzing textual and numerical data in
meta-analyses, but the system could not import information from
electronic databases \cite{ElaminM2009Choice}. 
Our original meta-analyses
\cite{KemptonM2011Structural,KemptonM2008MetaAnalysis} relied on the
Microsoft Excel spreadsheets later distributed on public web sites.
Compared to an ordinary spreadsheet a wiki solution provides
data entry provenance and collaborative data entry with immediately
update. 
Shareable folders on cloud-based storage systems would help
collaboration on spreadsheets, but yield no provenance. 
Online services, such as the spreadsheet of Google Docs, may lack
meta-analytic plotting facility.
Web-based specialized applications for systematic reviews may have a
high setup cost \cite{ElaminM2009Choice}. 

We have previously explored a simple online meta-analysis system---a
``fielded wiki''---in connection with personality genetics
\cite{NielsenF2010Fielded}. 
As implemented specifically for this scientific area the web service
lacks generality for other types for meta-analytic data. 
Furthermore the system relied on PubMed or Brede Wiki to represent
bibliographic information.

Following Ward Cunningham's quote ``What's the simplest thing that could
possibly work?''  
we present a simple system that allows for mass meta-analysis of
numerical data presented as comma-separated values (CSV) in a standard
Media\-Wiki-based wiki, --- the Brede Wiki: \url{http://neuro.imm.dtu.dk/wiki/}.

\section{Data and data representation}

We use the MediaWiki-based Brede Wiki to represent the data
\cite{NielsenF2009BredeWikiNeuroscience}. 
For our neuroimaging data each data record usually consists of three values
(number of subjects, their mean and standard deviation). 
The individual study typically compares two such data records, e.g.,
from a patient and a control group. 
We also record labels for the data record, e.g., the biographic
information, as well as extra subject information about the two
groups, such as age, gender and clinical characteristics, so that the
total number of data items for each study may be seven or more.
Each meta-analysis will usually determine what extra relevant
information should be included and it may differ between studies,
e.g., a \emph{Y-BOCS} value has typically only relevance for
obsessive-compulsive disorder patients.
The functional neuroimaging area has \emph{CogPO} and
\emph{Cognitive Atlas} ontologies enabling\linebreak researchers to describe
the topic of an experiment, but these efforts do not directly apply
to our data.
One CSV line carries the information for each study.

\begin{wrapfigure}[15]{r}{.6\textwidth}
  \begin{center}
    \vspace{-1.2cm}
    \includegraphics[width=.6\textwidth]{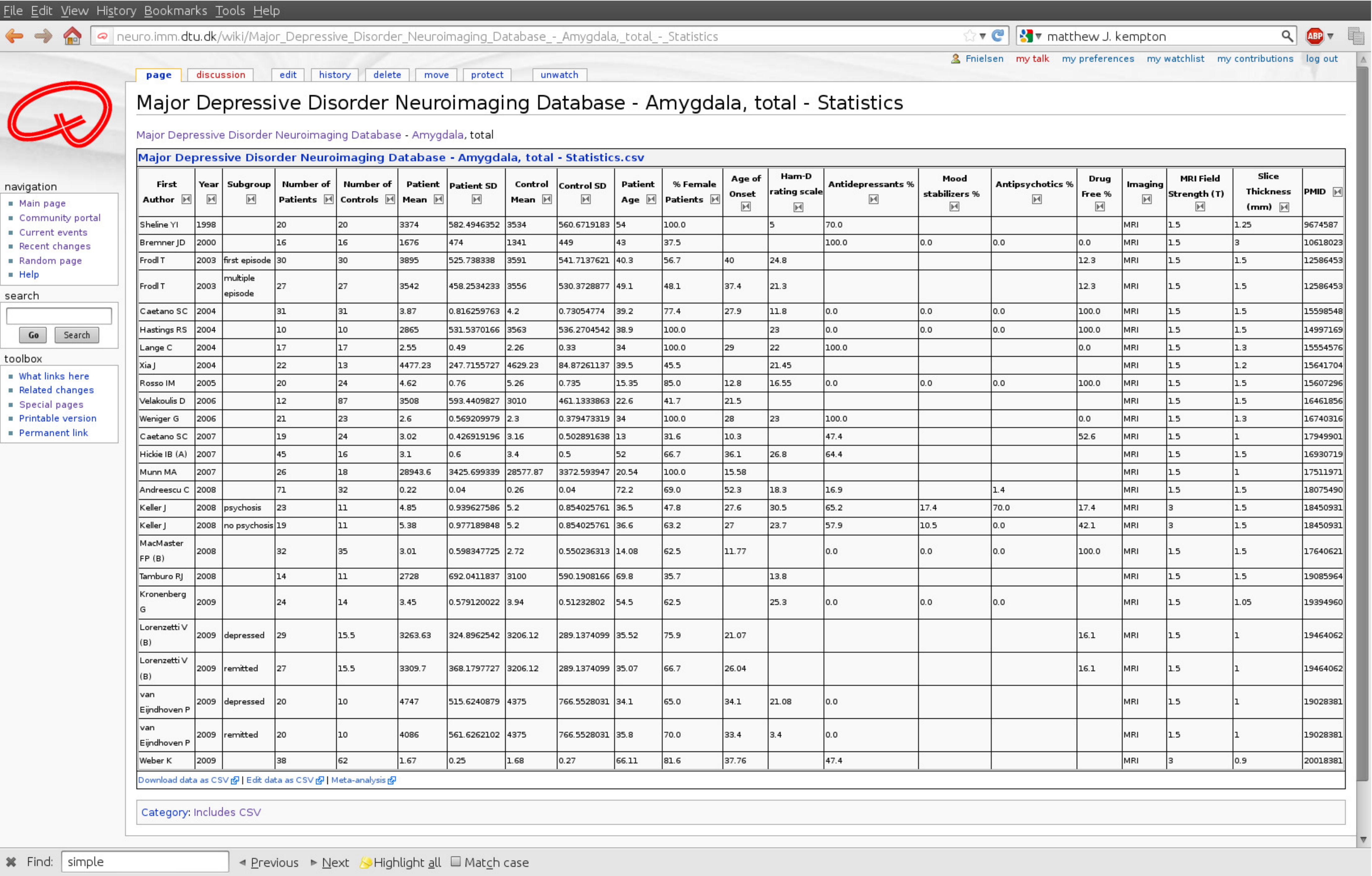}
    \vspace{-0.5cm}
    \caption{Screenshot from the wiki showing CSV data transcluded on
      a page.}
    \label{fig:datascreenshot}
  \end{center}
\end{wrapfigure}

Separate wiki pages store---rather than uploaded files---the CSV data,
so the Media\-Wiki template functionality can transclude the 
CSV data on other wiki pages.
By convention pages with CSV information have the ``.csv''
extension as 
part of the title so external scripts can recognize
them as special pages and the wiki pages have no wiki markup.

MediaWiki templates may generate links for download, editing and
meta-analysis of the data. 
Presently, no controlled vocabulary beyond the template fields
describes the columns in the CSV.
To generate an appropriate content-type (text/csv) a
bridging web script functions as a proxy, so a download of the CSV
page can spawn a client-side spreadsheet program.

A few MediaWiki extensions can format CSV information:
\emph{SimpleTable}
and \emph{TableData}.  
Figure~\ref{fig:datascreenshot} shows the transclusion of CSV data
with a modified version of the SimpleTable extension.
The Brede Wiki uses the standard template system for recording
structured bibliographic data about the publication and to annotate
the CSV information, see Figure~\ref{fig:templates}.

\begin{wrapfigure}[11]{r}{.6\textwidth}
  \vspace{-0.8cm}
  \begin{tabular}{|c|}
    \hline
    \begin{minipage}{\linewidth}
      {\tiny
\begin{verbatim}
{{Metaanalysis csv begin}}
{{Metaanalysis csv
 | title = Major Depressive Disorder Neuroimaging Database - Pituitary, total
 | topic1 = Pituitary
 | topic2 = Major depressive disorder
 | topic3 = MaND
}}
{{Metaanalysis csv
 | title = Obsessive-compulsive disorder Neuroimaging Database - Pituitary
 | topic1 = Pituitary
 | topic2 = Obsessive-compulsive disorder
 | topic3 = ObND
}}
{{Metaanalysis csv end}}
\end{verbatim}
}
\end{minipage}
\\
\hline
\end{tabular}
\caption{Template to annotate the CSV data and define the links to the meta-analysis.}
\label{fig:templates}
\end{wrapfigure}

The bulk of the data currently presented in the wiki comes from the
large mass meta-analysis of volumetric studies on major depressive
disorder reporting over 50 separate meta-analyses for individual
brain regions \cite{KemptonM2011Structural}.
Further data comes from mass
meta-analyses across multiple brain regions on bipolar disorder
\cite{KemptonM2008MetaAnalysis} and first-episode schizophrenia
\cite{SteenR2006Brain}, a meta-analysis on 
longitudinal development in schizophrenia
\cite{KemptonM2010Progressive} as well as 
data from individual original studies on obsessive-compulsive disorder.

Apart from neuroimaging studies the Brede Wiki also records data from
meta-analyses from a few other studies outside neuroimaging
\cite{HartungJ2008Statistical}, allowing us to test the generality of
the framework.
The data is distributed under ODbL.

\section{Web script and meta-analysis}

The web script for meta-analysis reads the CSV information, identifies
the required columns for meta-analysis, performs the statistical
computations and makes meta-analytic plots--- the so-called forest
and funnel plots---in the SVG format, see Figure~\ref{fig:screenshot}.
From either the title information or a PubMed identifier the script
generates back-links from the generated page to pages on the wiki.
The script may also export the computed results as JSON or CSV.
Furthermore, it may generate a small \emph{R} script that sets up
the data in variables and use the {\tt meta} library for meta-analysis.

The web script attempts to guess the separator used on the CSV page and
also tries to match the elements of the column header, e.g., the
strings ``control n'', ``controls number'', ``number of controls'',
etc.  match for number of control subjects.
With no matches the user needs to explicitly specify the relevant
columns via URL parameters, which in turn a wiki template can setup. 

Standard meta-analysis computes an \emph{effect size} from each result
in a paper and computes a combined meta-analytic effect size and its
confidence interval. 
Although the methodological development continues, there exist
established statistical analysis approaches for ordinary meta-analysis
\cite{HartungJ2008Statistical}.   
Our system implements computations on the standardized mean difference
for continuous variables and on the natural logarithm of the odds ratio
for categorical variables with fixed and random effects methods using
an inverse-weighted variance model, --- following the approach in the
Stata program. 
As an extra option we provide meta-analysis on the natural logarithm of
the variance ratio \cite{ShafferJ1992Caution}, for comparison of the
standard deviations between two groups of subjects.  


\section{Results}

\begin{wrapfigure}[20]{r}{.67\textwidth}
  \vspace{-.9cm}
  \centering
  \includegraphics[width=.67\textwidth]{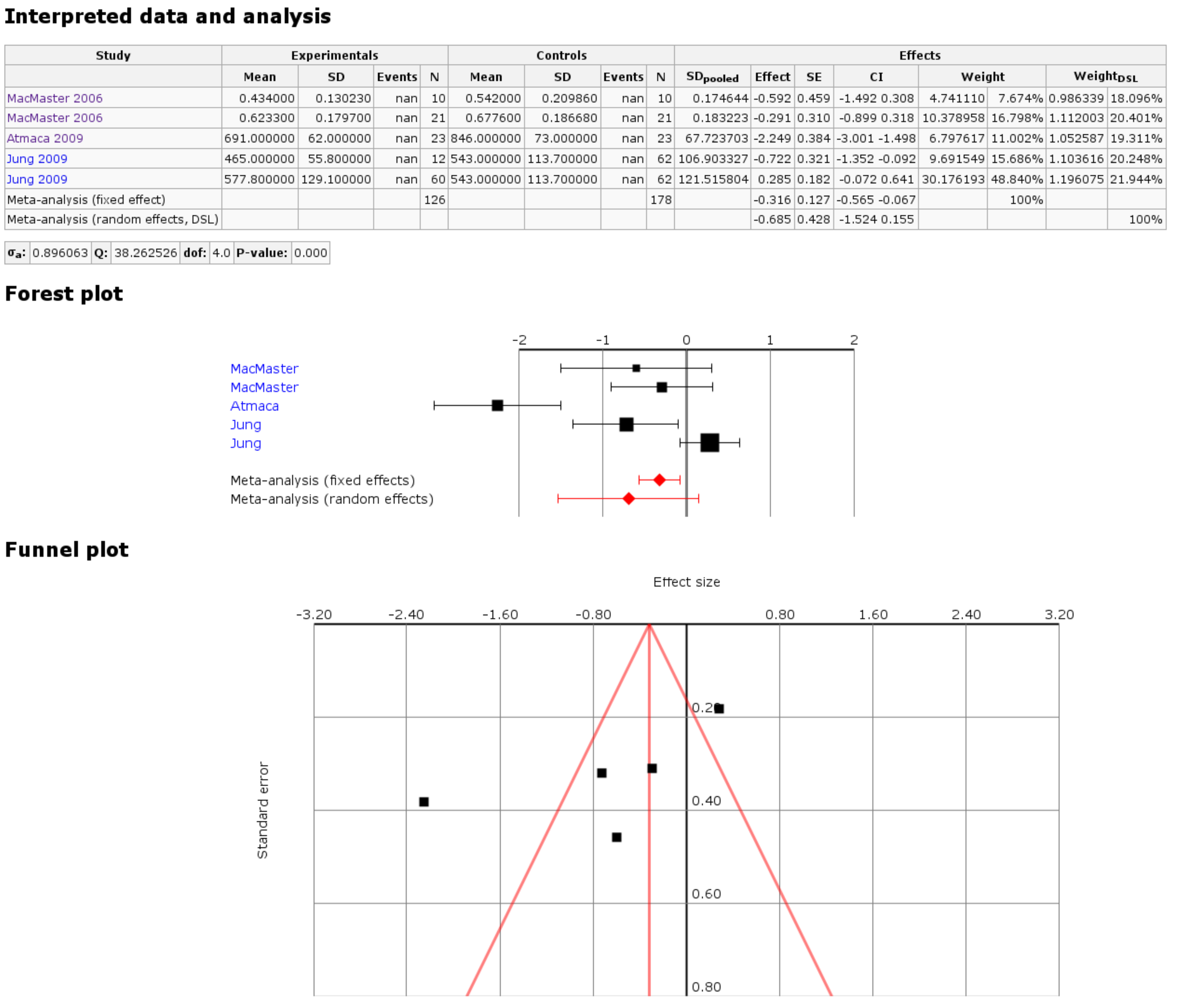}
  \caption{Screenshot of web script showing the meta-analytic results
    with forest and funnel plots.}
  \label{fig:screenshot}
\end{wrapfigure}

\begin{wrapfigure}[22]{r}{.68\textwidth}
  \vspace{-7mm}
  \hspace{-9mm} 
  \includegraphics[width=.79\textwidth]{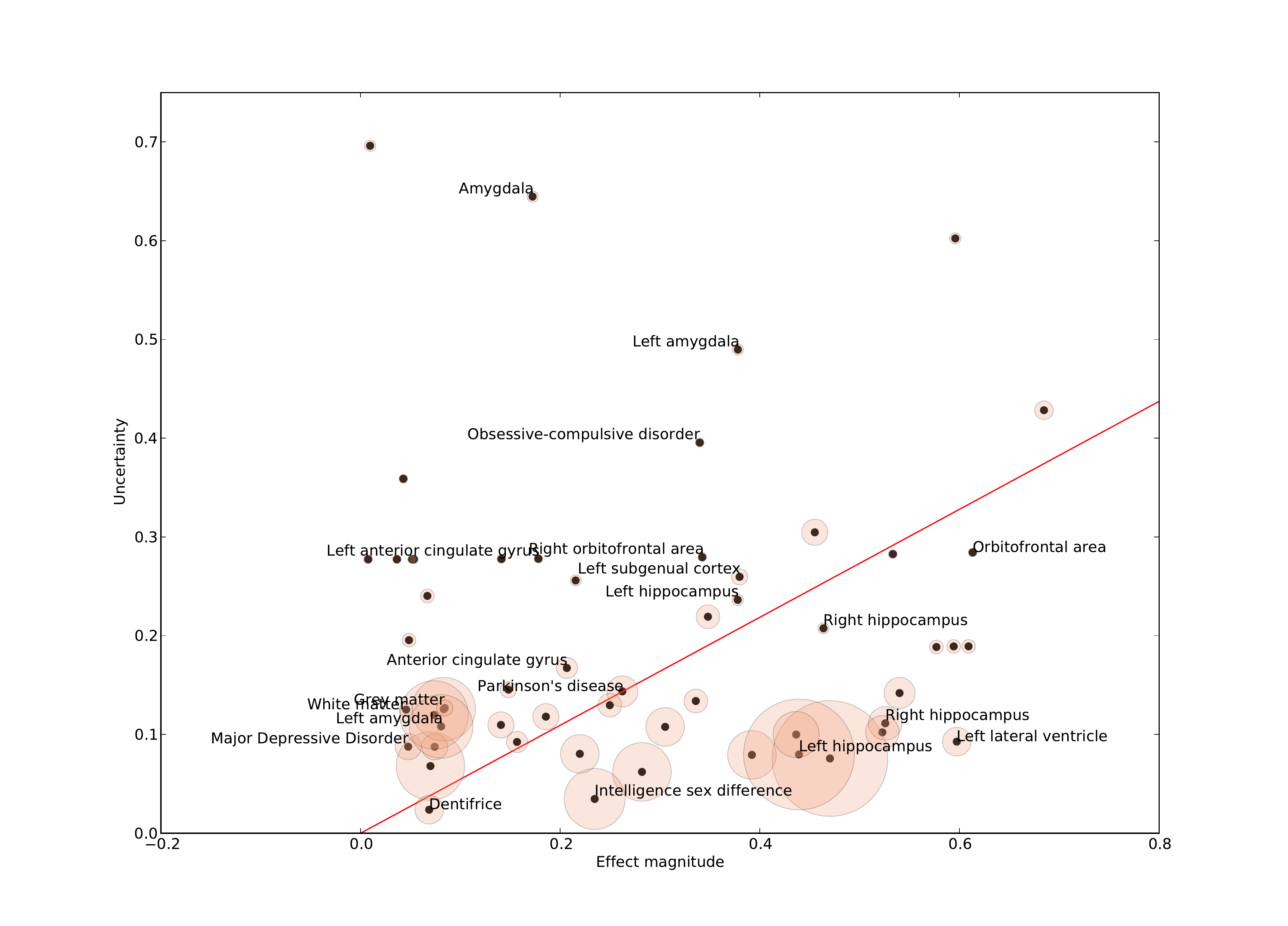}
  \vspace{-7mm}
  \caption{Results from mass meta-analyses shown in a L'Abbé-like
    plot and constructed by
    calling the web script  
    multiple times. Each dot corresponds to a
    meta-analysis. Uncertainty as a function of effect size with size
    of each dot determined by the number of subjects. The line
    indicates 0.05-significance.}  
  \label{fig:massmetaanalysis}
\end{wrapfigure}

We have added 124 pages with CSV data, --- most of which contain data
suitable for meta-analysis.
For individual analyses the reading, computation and download finish
within seconds. 
With multiple calls to the web script and JSON output another script
can plot multiple meta-analytic results together as in
Figure~\ref{fig:massmetaanalysis}. 
Generating such a plot takes several minutes.
For generating the page shown in Figure~\ref{fig:screenshot} we need
only the CSV data and 
the web script, while the script that generated
Figure~\ref{fig:massmetaanalysis} used information defined in
templates, CSV data and the web script with no further adaption of
MediaWiki. 

\section{Discussion}

By using MediaWiki in our present system we exploit the template facility
to capture structured information, and free-form wikitext for
annotation and comment on the individual scientific papers, --- as in
semantic academic annotation wikis \emph{AcaWiki} and 
  \emph{WikiPapers}. 
It is also possible to use the pages of the wiki as a simple means to keep track
of the status of the papers considered for the meta-analysis: potentially
eligible, eligible, partially entered and fully entered.

Why not Semantic MediaWiki?
Semantic MediaWiki (SMW) may query text and numerical data, though has not
had the ability to make complex computations. 
The \emph{Semantic Result
  Formats}
extension includes average, sum, 
product and count result formats enabling simple computations of
a series of numerical values, but insufficient for the kind of
computations we require. 
The data for meta-analysis form a n-ary data record (mean,
standard deviation, number of subjects, labels) so either individual
SMW pages should store each data record or we should invoke the n-ary
functionality in \emph{Semantic Internal
  Objects}
SMW extension, SMW \emph{record} or the recently-introduced
\emph{subobject} SMW functionality.
We have not investigated whether these tools provide convenient means for representing our data.
The Brede Wiki can export its ontologies defined in MediaWiki template
to SKOS.
Our future research can consider RDFication of the CSV information
through the SCOVO format \cite{HausenblasM2009SCOVO_without_editor}.


\sloppy

We wrote the web service in Python, where Numpy makes vector
computation available and Scipy provides statistical methods,
necessary for the computation. In a future PHP implementation the
script could more closely integrate 
with the wiki as either a Media\-Wiki or a SMW extension. 

A wiki built from standard components provides a
inexpensive solution with means to manage meta-analytic data in a
collaborative environment. 
The general framework allows not only the meta-analysis of
neuroimaging-derived data but has the potential for managing and analyzing
data from many other domains.

\bibliographystyle{splncs}

\end{document}